\documentclass{revtex4}
\usepackage{graphicx}
\usepackage{amsmath}
\usepackage{amssymb}
\usepackage{float}
\usepackage{color}

\begin{document}
\title{
Scalar model of SU(N) glueball \`{a} la Heisenberg
}
\author{
Vladimir Dzhunushaliev
}
\email{v.dzhunushaliev@gmail.com}
\affiliation{
Dept. Theor. and Nucl. Phys., KazNU, Almaty, 050040, Kazakhstan
}
\affiliation{
IETP, Al-Farabi KazNU, Almaty, 050040, Kazakhstan
}

\author{Vladimir Folomeev}
\email{vfolomeev@mail.ru}
\affiliation{Institute of Physicotechnical Problems and Material Science of the NAS
of the Kyrgyz Republic, 265 a, Chui Street, Bishkek 720071,  Kyrgyz Republic}

\begin{abstract}
Nonperturbative model of glueball is studied. The model is based on the nonperturbative quantization technique suggested by Heisenberg.
2- and 4-point Green functions for a gauge potential are expressed in terms of two scalar fields. The first scalar field describes quantum fluctuations
of the subgroup $SU(n) \subset SU(N)$, and the second one describes quantum fluctuations of the coset $SU(N) / SU(n)$.
An effective Lagrangian for the scalar fields is obtained.
The coefficients for all terms in the Lagrangian are calculated, and it is shown that they depend on
$\dim SU(n), \dim SU(N)$. It is demonstrated that a spherically symmetric solution describing the glueball does exist.
\end{abstract}

\maketitle

\section{Introduction}

Glueball is a hypothetical particle appearing  in SU(3) quantum gauge field theory only.
Roughly speaking, one can say that the glueball is a proton (or a neutron) with remote quarks.
The existence of glueballs is a consequence of the self-interaction of gluons within QCD.
Fritzsch and Minkowski~\cite{Fritzsch:1975tx} developed a model of glueball
on the basis of the quark-gluon field theory by assuming an analogy between massless gluons and photons.
Jaffe and Johnson~\cite{Jaffe:1975fd} investigated the bag model of glueball.
At the present time, several candidates for low-mass glueballs with quantum numbers $0^{++}, 2^{++}, 0^{- +}$ and $1^{- -}$
are under discussion~\cite{Klempt:2007cp,Ochs:2006rb,Ochs:2013gi,Mathieu:2008me}.

Historically, the idea of studying QCD in the large-N limit was first put forward in 1974 by 'tHooft~\cite{'tHooft:1973jz},
who proposed to consider $1/N$ as an expansion parameter. Some nonperturbative features of QCD can be understood in such a large-$N$  limit~\cite{Witten:1979kh,Manohar:1998xv}.

Here we would like to calculate the mass of a glueball in SU(N) gauge theory by using the nonperturbative quantization technique \`{a} la Heisenberg \cite{heis}.
The essence of such a technique consists in
writing equation(s) for field operator(s). In fact, such equation(s) is(are)
equivalent to an infinite set of equations  for all Green functions~\cite{Dzhunushaliev:2015hoa}. Mathematically, such procedure is similar to writing down of infinite
system of equations for all cumulants in turbulence modeling \cite{Landau:1982dva,Wilcox}. Since it is impossible to solve such an infinite set of equations,
one can proceed similarly to
the case of turbulence modeling  when one has to cut off the set of equations using some assumptions concerning Green functions (the closure problem).
For example, this can be done by introducing some approximate connection between higher- and lower-order Green functions. In addition, we will use the following assumptions:
(a) 2-point Green functions of SU(N) gauge potentials can approximately be expressed through scalar functions; (b) 2-point Green functions
for the subgroup SU(n) and the coset SU(N)/SU(n) are different. Namely, we employ $\chi$ for SU(n) 2,4-point Green functions and $\phi$ for SU(N)/SU(n)
coset 2,4-point Green functions; (c) there is an anisotropy in the color space in the sense that Green functions are not symmetric under exchange of color indices.

Based on the above assumptions and ideas, we have the following approximate model of a glueball: (a) it is a ball filled with fluctuating SU(N) quantum gauge fields;
(b) quantum fields are approximately described by two scalar fields; (c) one of the fields describes the fluctuating SU(n) components, and another one -- the coset SU(N)/SU(n)
components; (d) 2- and 4-point Green functions are nonsymmetric in the color space under exchange of color indices; (e) the approximate description
of the glueball is therefore carried out by using nonlinear equations for two scalar fields; (f) the glueball is described by spherically symmetric solutions of these equations;
(g) using the expression for the scalar fields energy density, we can calculate the mass of the glueball within this model;
(h) all components of the subgroup SU(n) give similar contributions to a Green function;  the same holds true for the coset SU(N)/SU(n).

It will be shown below that
equations for the scalar fields contain  parameters $\lambda_{1,2,3}$, describing the self-interaction of the scalar fields (gauge potentials), and
the value $\chi(0)$ of one of the scalar fields at the origin (dispersion of fluctuations of the gauge potentials). After fixing $\lambda_{1,2,3}$,
one can show that the dimensionless glueball mass for SU(N) gauge field theory depends on the value of one scalar field at the origin
and the dimensions of the subgroup SU(n) and the gauge group SU(N):
$n = \dim SU(n), N = \dim SU(N)$.

\section{Scalar model of a glueball}

Our approximation is based on the main assumption that 2- and 4-points Green functions are described in terms of some scalar fields $\phi$ and $\chi$
according to the following relations (which are some variations of ans\"atz from \cite{Dzhunushaliev:2015hoa}):
\begin{eqnarray}
	\left( G_2 \right)^{ab}_{\mu \nu}(x, x) & = &
	\left\langle
		A^a_\mu (x) A^b_\nu (x)
	\right\rangle
	\approx
	C^{ab} \tilde b_{\mu \nu} {\tilde \chi}^2(x)  ,
\label{1-10}\\
	\left( G_2 \right)^{ab}_{\mu \nu; \alpha \beta}(x, x) & = &
	\left\langle
		\partial_\alpha A^a_\mu (x) \partial_\beta A^b_\nu (x)
	\right\rangle
	\approx
	C^{ab} \tilde b_{\mu \nu} \partial_\alpha \tilde \chi(x) \partial_\beta \tilde \chi(x) ,
\label{1-20}\\
	\left( G_4 \right)^{abcd}_{\mu \nu \rho \sigma}(x, x, x, x) &=&
	\left\langle
		A^a_\mu (x) A^b_\nu (x) A^c_\rho (x) A^d_\sigma (x)
	\right\rangle
	\approx
\nonumber \\
  &&
	\Bigl ( \left\langle
		A^a_\mu (x) A^b_\nu (x)
	\right\rangle - m^{ab}_{\mu \nu} \Bigl )
	\Bigl ( \left\langle
		A^c_\rho (x) A^d_\sigma (x)
	\right\rangle - m^{cd}_{\rho \sigma} \Bigl ) -
  m^{ab}_{\mu \nu} m^{cd}_{\rho \sigma} =
\nonumber \\
  &&
  C^{ab} \tilde b_{\mu \nu} C^{cd} \tilde b_{\rho \sigma}  \left[ \left(
    \tilde \chi^2(x) - \tilde m_2^2
  \right)^2 - \tilde m_2^4 \right] ,
\label{1-30}\\
	\left( G_2 \right)^{mn}_{\mu \nu}(x, x) &=&
	\left\langle
		A^m_\mu (x) A^n_\nu (x)
	\right\rangle
	\approx
	C^{mn} \tilde{\tilde b}_{\mu \nu} \left(
    \tilde m_1^2 - \tilde \phi^2(x)
  \right) ,
\label{1-40}\\
	\left( G_2 \right)^{mn}_{\mu \nu; \alpha \beta}(x, x) & = &
	\left\langle
		\partial_\alpha A^m_\mu (x) \partial_\beta A^n_\nu (x)
	\right\rangle
	\approx
	C^{mn} \tilde{\tilde b}_{\mu \nu}
  \partial_\alpha \tilde \phi(x) \partial_\beta \tilde \phi(x) ,
\label{1-45}\\
	\left( G_4 \right)^{mnpq}_{\mu \nu \rho \sigma}(x, x, x, x) &=&
	\left\langle
		A^m_\mu (x) A^n_\nu (x) A^p_\rho (x) A^q_\sigma (x)
	\right\rangle
	\approx
	\left\langle
		A^m_\mu (x) A^n_\nu (x)
	\right\rangle
	\left\langle
		A^p_\rho (x) A^q_\sigma (x)
	\right\rangle =
\nonumber \\
  &&
  C^{mn} \tilde{\tilde b}_{\mu \nu} C^{pq} \tilde{\tilde b}_{\rho \sigma} \left(
    \tilde \phi^2(x) - \tilde m_1^2
  \right)^2 ,
\label{1-50} \\
	\left( G_4 \right)^{ab mn}_{\mu \nu \rho \sigma}(x, x, x, x) &=&
	\left\langle
		A^a_\mu (x) A^b_\nu (x) A^m_\rho (x) A^n_\sigma (x)
	\right\rangle
	\approx
	\left\langle
		A^a_\mu (x) A^b_\nu (x)
	\right\rangle
	\left\langle
		- A^m_\rho (x) A^n_\sigma (x) + C^{mn}_{\mu \nu} m_1^2
	\right\rangle =
\nonumber \\
  &&
  C^{ab} \tilde b_{\mu \nu} C^{mn} \tilde{\tilde b}_{\rho \sigma}
    {\tilde \phi}^2(x) \tilde \chi^2(x),
\label{1-60}
\end{eqnarray}
where $a,b,c,d = 1,2, \dots , n$ are the SU(n) indices, $m,n,p,q = 4,5, \cdots , N$ are the coset $SU(N) / SU(n)$ indices,
$C^{ab, mn}, \tilde b_{\mu\nu}, \tilde{\tilde b}_{\mu \nu}$, and $m^{ab, mn}_{\mu\nu}$ are the closure constants.
We see that, similarly to turbulence modeling, we have to introduce some closure constants.

\section{Estimation of numerical factors depending on
$\dim SU(N), \dim SU(n)$}

To begin with, let us consider the SU(N) Lagrangian
\begin{equation}
	- \mathcal L_{SU(N)}  =  \frac{1}{4} F^B_{\mu \nu} F^{B \mu \nu} =
  \frac{1}{4} F^a_{\mu \nu} F^{a \mu \nu} +
  \frac{1}{4} F^m_{\mu \nu} F^{m \mu \nu} ,
\label{2-10}
\end{equation}
where $B = 1,2, \ldots , \dim SU(N)$ is the SU(N) index,
$F^B_{\mu \nu} = \partial_\mu A^B_\nu - \partial_\nu A^B_\mu +
g f^{BCD} A^C_\mu A^D_\nu$ is the field strength operator, $g$ is the coupling constant, and
\begin{eqnarray}
  F^a_{\mu \nu} &=& f^a_{\mu \nu} f^{a\mu \nu} +
  g f^{abc} A^b_\mu A^c_\nu + g f^{amn} A^m_\mu A^n_\nu ,
\label{2-20}\\
  F^m_{\mu \nu} &=& f^m_{\mu \nu} f^{m \mu \nu} +
	g f^{mpq} A^{p}_\mu A^{q}_\nu +
	g f^{mna} \left(
    A^{n}_\mu A^{a}_\nu - A^{n}_\nu A^{a}_\mu
  \right).
\label{2-30}
\end{eqnarray}
Here $f^B_{\mu \nu} = \partial_\mu A^B_\nu -\partial_\nu A^B_\mu $;
$a, b, c = 1,2, \ldots , n$ are the SU(n) indices; $m, n, p, q$ are the coset indices $\in SU(N)/SU(n)$.
In order to obtain an effective Lagrangian, we have to average the SU(N) Lagrangian \eqref{2-10} over a quantum state $\left. \left. \right| Q \right\rangle$.
In this case we will have the following expressions for both terms from \eqref{2-10} :
\begin{eqnarray}
  \left\langle
    F^a_{\mu \nu} F^{a \mu \nu}
  \right\rangle &=&
  \left\langle
    f^a_{\mu \nu} f^{a\mu \nu}
  \right\rangle +
  g^2 f^{a b_1 c_1} f^{a b_2 c_2}
  \left\langle
    A^{b_1}_\mu A^{c_1}_\nu A^{b_2 \mu} A^{c_2 \nu}
  \right\rangle
  + g^2 f^{a m_1 n_1} f^{a m_2 n_2}
  \left\langle
    A^{m_1}_\mu A^{n_1}_\nu A^{m_2 \mu} A^{n_2 \nu}
    \right\rangle +
\nonumber \\
  &&
  g^2 f^{a b c} f^{a m n}
  \left\langle
    A^{b}_\mu A^{c}_\nu A^{m \mu} A^{n \nu}
  \right\rangle ,
\label{2-40}\\
  \left\langle
  F^m_{\mu \nu} F^{m \mu \nu}
  \right\rangle &=&
  \left\langle
    f^m_{\mu \nu} f^{m \mu \nu}
  \right\rangle +
	g^2 f^{m p_1 q_1} f^{m p_2 q_2}
  \left\langle
    A^{p_1}_\mu A^{q_1}_\nu A^{p_2 \mu} A^{q_2 \nu}
  \right\rangle +
	g^2 f^{m p_1 a_1} f^{m p_2 a_2} \Bigl(
  \left \langle
    A^{p_1}_\mu A^{a_1}_\nu A^{p_2 \mu} A^{a_2 \nu}
  \right\rangle +
\nonumber \\
  &&
  \left\langle
    A^{a_1}_\nu A^{p_1}_\mu A^{a_2 \nu} A^{p_2 \mu}
  \right\rangle -
  \left\langle
    A^{p_1}_\mu A^{a_1}_\nu A^{a_2 \nu} A^{p_2 \mu}
  \right\rangle -
  \left\langle
    A^{a_1}_\nu A^{p_1}_\mu A^{p_2 \mu} A^{a_2 \nu}
  \right\rangle
  \Bigl).
\label{2-50}
\end{eqnarray}
Here we assume that all terms having odd number of potential components are zero:
$
\left\langle A^{a,m}_\mu \right\rangle =
\left\langle A^{a,m}_\mu A^{b,n}_\nu A^{c,p}_\rho \right\rangle = 0
$ and
$
\left\langle A^{a}_\mu A^{m}_\nu A^{n}_\nu \right\rangle \approx
\left\langle A^{a}_\mu  \right\rangle \left\langle A^{m}_\nu A^{n}_\nu \right\rangle
= 0$.

Our goal is to calculate the effective Lagrangian
\begin{equation}
	\mathcal L_{\rm{eff}} \approx \left\langle \mathcal L_{SU(N)} \right\rangle .
\label{2-60}
\end{equation}
To do this, let us consider first terms on the right-hand sides of Eqs.~\eqref{2-40} and \eqref{2-50}:
\begin{equation}\label{2-70}
  \left\langle
    f^a_{\mu \nu} f^{a\mu \nu}
  \right\rangle = 2 \left\langle
    \partial_\mu A^a_\nu \partial^\mu A^{a \nu}
  \right\rangle - 2 \left\langle
    \partial_\mu A^a_\nu \partial^\nu A^{a \mu}
  \right\rangle .
\end{equation}
Substituting here the expression \eqref{1-10}, we obtain
\begin{equation}\label{2-75}
\begin{split}
  &\left\langle
    \partial_\mu A^a_\nu \partial^\mu A^{a \nu}
  \right\rangle - \left\langle
    \partial_\mu A^a_\nu \partial^\nu A^{a \mu}
  \right\rangle =
  C^{aa} \left(
    b^\alpha_\alpha \eta^{\mu \nu} - b^{\mu \nu}
  \right) \partial_\mu \chi \partial_\nu \chi =
\\
  &
  C^{aa} \Bigl[
    - \left( b^{11} + b^{22} + b^{33} \right)\partial_0 \chi \partial_0 \chi +
    \left( - b^{00} + b^{22} + b^{33} \right)\partial_1 \chi \partial_1 \chi +
    \left( - b^{00} + b^{11} + b^{33} \right)\partial_2 \chi \partial_2 \chi +
\\
    &
    \left( - b^{00} + b^{11} + b^{22} \right)\partial_3 \chi \partial_3 \chi -
    \sum\limits_{\mu \neq \nu} b^{\mu \nu} \partial_\mu \chi \partial_\nu \chi
  \Bigl].
\end{split}
\end{equation}
It is seen that the structure of the term
$\left\langle f^a_{\mu \nu} f^{a\mu \nu} \right\rangle$ is very confusing. As a toy model,
consider the case $b_{\mu \nu} = - \eta_{\mu \nu}$, where $\eta_{\mu \nu}$ is Minkowski metric.
Let us note that
in this case $\left\langle \left( A^a_0 \right)^2 \right\rangle < 0$. This is the price for using such an approximation.

As a first approximation we assume that all numbers $C^{ab}$ in Eq.~\eqref{1-20} have the same order, and this factor can be estimated as
$C^{ab} \approx C_2$. In this case \eqref{2-70} has the form
\begin{equation}\label{2-80}
  \left\langle
    f^a_{\mu \nu} f^{a\mu \nu}
  \right\rangle \approx - 6 C_2
  \left( n^2 - 1 \right)
  \partial_\mu \tilde \chi \partial^\mu \tilde \chi.
\end{equation}
Here we took into account that the summation over $a$ in $\sum_a^{\dim SU(n)}$ has $\left(n^2 - 1\right)$ summands,
$C^{aa} \approx C_2 \left( n^2 - 1 \right)$, and $C_2$ is a constant describing the dispersion of one component of the gauge potential $A^a_\mu$.
Using the same approximation, we obtain
\begin{equation}\label{2-90}
  \left\langle f^m_{\mu \nu} f^{m \mu \nu} \right\rangle \approx - 6 C_1
  \left( N^2 - n^2 \right)
  \partial_\mu \tilde \phi \partial^\mu \tilde \phi.
\end{equation}
Here we took into account that the summation over index $m$ has
$\dim SU(N) - \dim SU(n)= N^2 - n^2$ summands,
$C^{mm} = C_1 \left( N^2 - n^2 \right)$, and $C_1$ is a constant describing the dispersion of one component of the gauge potential $A^m_\mu$.

We assume that our physical situations can be described by the ansatz \eqref{1-30} for the 4-point Green function, and consequently
\begin{equation}\label{2-110}
  f^{a b_1 c_1} f^{a b_2 c_2}
  \left\langle
    A^{b_1}_\mu A^{c_1}_\nu A^{b_2 \mu} A^{c_2 \nu}
    \right\rangle \approx
  f^{a b_1 c_1} C^{b_1 c_1} f^{a b_2 c_2} C^{b_2 c_2}
  \left[
    \left( \chi^2 - \tilde m_2^2 \right)^2 - \tilde m_2^4
  \right] , \quad b_{1,2} \neq c_{1,2}.
\end{equation}
Let us consider the summands with fixed $a$ in more detail:
\begin{equation}\label{2-115}
\begin{split}
  & f^{a b_1 c_1} f^{a b_2 c_2} C^{b_1 c_1} C^{b_2 c_2} +
  f^{a b_1 c_1} f^{a c_2 b_2} C^{b_1 c_1} C^{c_2 b_2} +
  f^{a c_1 b_1} f^{a b_2 c_2} C^{c_1 b_1} C^{b_2 c_2} +
\\
  &
  f^{a c_1 b_1} f^{a c_2 b_2} C^{c_1 b_1} C^{c_2 b_2} =
   f^{a b_1 c_1} f^{a b_2 c_2}
  \left( C^{b_1 c_1} - C^{c_1 b_1} \right)
  \left( C^{b_2 c_2} - C^{c_2 b_2} \right) =
  f^{a b_1 c_1} f^{a b_2 c_2} \Delta C^{b_1 c_1} \Delta C^{b_2 c_2}
\end{split}
\end{equation}
(no summation over repeated  indices). The number of such terms is equal to the number of pairs of the structure
constants $f^{a b_1 c_1}, f^{a b_2 c_2} $ with $a, b_{1,2}, c_{1,2} \in SU(n)$ and different $a$.
Now we want to estimate the term from \eqref{2-110} as
\begin{equation}\label{2-100}
  	f^{a b_1 c_1} f^{a b_2 c_2}
  \left\langle
    A^{b_1}_\mu A^{c_1}_\nu A^{b_2 \mu} A^{c_2 \nu}
  \right\rangle \approx \frac{
    \Bigl( \tilde \lambda_2 \Bigl)_{n, n, n}^{n, n, n}}{4}
  \left[
    \left( \tilde \chi^2 - \tilde m_2^2 \right)^2 - \tilde m_2^4
  \right],
\end{equation}
where
$\Bigl( \tilde \lambda_2 \Bigl)_{n, n, n}^{n, n, n} =
4 f^{a b_1 c_1} f^{a b_2 c_2} \Delta C^{b_1 c_1} \Delta C^{b_2 c_2}$.

Next term is
\begin{equation}\label{2-112}
  f^{a m_1 n_1} f^{a m_2 n_2}
  \left\langle
    A^{m_1}_\mu A^{n_1}_\nu A^{m_2 \mu} A^{n_2 \nu}
    \right\rangle \approx
  f^{a m_1 n_1} C^{m_1 n_1} f^{a m_2 n_2} C^{m_2 n_2}
    \left( \tilde \phi^2 - m_1^2 \right)^2  , m_{1,2} \neq n_{1,2} .
\end{equation}
Proceeding as in \eqref{2-110}, we have
\begin{equation}\label{2-114}
\begin{split}
  & f^{a m_1 n_1} f^{a m_2 n_2} C^{m_1 n_1} C^{m_2 n_2} +
  f^{a m_1 n_1} f^{a n_2 m_2} C^{m_1 n_1} C^{n_2 m_2} +
  f^{a n_1 m_1} f^{a m_2 n_2} C^{n_1 m_1} C^{m_2 n_2} +
\\
  &
  f^{a n_1 m_1} f^{a n_2 m_2} C^{n_1 m_1} C^{n_2 m_2} =
   f^{a m_1 n_1} f^{a m_2 n_2}
  \left( C^{m_1 n_1} - C^{n_1 m_1} \right)
  \left( C^{m_2 n_2} - C^{n_2 m_2} \right) =
\\
  &
  f^{a m_1 n_1} f^{a m_2 n_2} \Delta C^{m_1 n_1} \Delta C^{m_2 n_2}
\end{split}
\end{equation}
(again no summation over repeated indices). Now we can estimate this term as
\begin{equation}\label{2-102}
  	f^{a m_1 n_1} f^{a m_2 n_2}
  \left\langle
    A^{m_1}_\mu A^{n_1}_\nu A^{m_2 \mu} A^{n_2 \nu}
  \right\rangle \approx
  \frac{\Bigl( \tilde \lambda_1 \Bigl)_{n, N/n, N/n}^{n, N/n, N/n}}{4}
  \left( \tilde \phi^2 - \tilde m_1^2 \right)^4,
\end{equation}
where
$\Bigl( \tilde \lambda_1 \Bigl)_{n, N/n, N/n}^{n, N/n, N/n} =
4 f^{a m_1 n_1} f^{a m_2 n_2} \Delta C^{m_1 n_1} \Delta C^{m_2 n_2}$.

Next term in Eq.~\eqref{2-40} is
\begin{equation}\label{2-150}
  f^{a b c} f^{a m n}
  \left\langle
    A^{b}_\mu A^{c}_\nu A^{m \mu} A^{n \nu}
  \right\rangle \approx  f^{a b c}
  \left\langle
    A^{b}_\mu A^{c}_\nu
  \right\rangle
  f^{a m n} \left\langle
     A^{m \mu} A^{n \nu}
  \right\rangle =
  f^{a b c} f^{a m n} C^{bc} C^{mn} \tilde \phi^2 \tilde \chi^2,
\end{equation}
for which we have
\begin{equation}\label{2-155}
\begin{split}
  & f^{a bc} f^{a mn} C^{ab} C^{mn} +
  f^{a bc} f^{a nm} C^{ab} C^{nm} +
  f^{a cb} f^{a mn} C^{cb} C^{mn} +
\\
  &
  f^{a cb} f^{a nm} C^{cb} C^{nm} =
   f^{a bc} f^{a mn}
  \left( C^{bc} - C^{cb} \right)
  \left( C^{mn} - C^{nm} \right) =
  f^{a bc} f^{a mn} \Delta C^{ab} \Delta C^{mn}.
\end{split}
\end{equation}
Consequently,
\begin{equation}\label{2-157}
  f^{a b c} f^{a m n}
  \left\langle
    A^{b}_\mu A^{c}_\nu A^{m \mu} A^{n \nu}
  \right\rangle \approx
  \frac{\Bigl( \tilde \lambda_3 \Bigl)_{n, n, n}^{n, N/n, N/n}}{2}
  \tilde \phi^2 \tilde \chi^2,
\end{equation}
where
$\Bigl( \tilde \lambda_3 \Bigl)_{n, n, n}^{n, N/n, N/n} =
2 f^{a bc} f^{a mn} \Delta C^{ab} \Delta C^{mn}$.

Similarly,  we can estimate the second term in Eq.~\eqref{2-50} as follows:
\begin{equation}\label{2-120}
  	f^{m p_1 q_1} f^{m p_2 q_2}
  \left\langle
    A^{p_1}_\mu A^{q_1}_\nu A^{p_2 \mu} A^{q_2 \nu}
  \right\rangle \approx
  f^{m p_1 q_1} f^{m p_2 q_2} C^{p_1 q_1} C^{p_2 q_2}
  \left[
    \left( \tilde \phi^2 - \tilde m_1^2 \right)^2
  \right], p_{1,2} \neq q_{1,2} .
\end{equation}
Proceeding as in \eqref{2-110}, we have
\begin{equation}\label{2-125}
\begin{split}
  & f^{m p_1 q_1} f^{m p_2 q_2} C^{p_1 q_1} C^{p_2 q_2} +
  f^{m p_1 q_1} f^{m q_2 p_2} C^{p_1 q_1} C^{q_2 p_2} +
  f^{m q_1 p_1} f^{m p_2 q_2 C^{q_1 p_1} C^{p_2 q_2}} +
\\
  &
  f^{m q_1 p_1} f^{m q_2 p_2} C^{q_1 p_1} C^{q_2 p_2} =
   f^{m p_1 q_1} f^{m p_2 q_2}
  \left( C^{p_1 q_1} - C^{q_1 p_1} \right)
  \left( C^{p_2 q_2} - C^{q_2 p_2} \right) =
  f^{m p_1 q_1} f^{m p_2 q_2} \Delta C^{p_1 q_1} \Delta C^{p_2 q_2}
\end{split}
\end{equation}
(no summation over indices). Then
\begin{equation}\label{2-127}
  f^{m p_1 q_1} f^{m p_2 q_2}
  \left\langle
    A^{p_1}_\mu A^{q_1}_\nu A^{p_2 \mu} A^{q_2 \nu}
  \right\rangle \approx
    \frac{\Bigl( \tilde {\tilde \lambda}_1 \Bigl)_{N/n, N/n, N/n}^{N/n, N/n, N/n}}{4}
    \left( \tilde \phi^2 - \tilde m_1^2 \right)^2,
\end{equation}
where
$\Bigl( \tilde \lambda_1 \Bigl)_{N/n, N/n, N/n}^{N/n, N/n, N/n} =
4 f^{m p_1 q_1} f^{m p_2 q_2} \Delta C^{p_1 q_1} \Delta C^{p_2 q_2}$.

Finally, the last term in Eq.~\eqref{2-50} can be estimated as
\begin{equation}\label{2-130}
  f^{m p_1 a_1} f^{m p_2 a_2} \left\langle
  A^{p_1}_\mu A^{a_1}_\nu A^{p_2 \mu} A^{a_2 \nu} \right\rangle
  \approx
  f^{m p_1 a_1} f^{m p_2 a_2}
  \left\langle
  A^{a_1}_\nu A^{a_2 \nu} \right\rangle
  \left\langle
  A^{p_1}_\mu A^{p_2 \mu} \right\rangle =
  f^{m p_1 a_1} f^{m p_2 a_2} C^{p_1 a_1} C^{p_2 a_2} \tilde \phi^2 \tilde \chi^2.
\end{equation}

Let us now consider the summands with fixed $m$ in more detail:
\begin{equation}\label{2-135}
  C^{a_1 a_2} C^{p_1 p_2} \left(
    f^{m p_1 a_1} f^{m p_2 a_2} +
    f^{m p_1 a_1} f^{m a_2 p_2} +
    f^{m a_1 p_1} f^{m p_2 a_2} +
    f^{m a_1 p_1} f^{m a_2 p_2}
  \right) = 0
\end{equation}
(no summation over indices). Consequently, in our approximation
\begin{equation}\label{2-137}
  f^{m p_1 a_1} f^{m p_2 a_2} \left\langle
  A^{p_1}_\mu A^{a_1}_\nu A^{p_2 \mu} A^{a_2 \nu} \right\rangle  = 0.
\end{equation}

Notice that
in Eqs.~\eqref{2-110}-\eqref{2-130} we have used the assumption that
$C^{AB}$ [where $A,B \in SU(N)$] are not symmetric, i.e., $C^{AB} - C^{BA} \neq 0$.

Combining all expressions for 2- and 4-point Green functions, we obtain the following approximate Lagrangian:
\begin{equation}\label{2-140}
  \mathcal L_{eff} = \frac{1}{2} \left( N^2 - n^2 \right)
  \partial_\mu \bar \phi \partial^\mu \bar\phi +
  \frac{1}{2}  \left( n^2 - 1 \right)
  \partial_\mu \bar\chi \partial^\mu \bar\chi -
  \frac{\bar\lambda_1}{4}
   \left( \bar{\phi}^2 - \bar m_1^2 \right)^2 -
  \frac{\bar\lambda_2}{4}
  \left[
    \left( \bar\chi^2 -\bar m_2^2 \right)^2 -\bar m_2^4
  \right] -
  \frac{\bar\lambda_3}{2}
 \bar \phi^2 \bar \chi^2,
\end{equation}
where the coefficients $C_{1,2}$ have been eliminated by redefining
$\tilde\chi, \tilde\phi$ and $\tilde m_{1,2}$:
\begin{eqnarray}
 \bar \phi &=& \tilde \phi \sqrt{12 C_1}  ,
\label{2-152} \\
 \bar \chi &=& \tilde \chi \sqrt{12 C_2}  ,
\label{2-154} \\
\bar  m_1 &=& \tilde m_1 \sqrt{12 C_1}  ,
\label{2-156} \\
\bar  m_2 &=& \tilde m_2 \sqrt{12 C_2}.
\label{2-158}
\end{eqnarray}
Also, for brevity, we have introduced 
\begin{eqnarray}
  \bar \lambda_1 &=& \frac{
  \Bigl( \tilde \lambda_1 \Bigl)_{n, N/n, N/n}^{n, N/n, N/n} +
  \Bigl( \tilde{\tilde \lambda}_1 \Bigl)_{N/n, N/n, N/n}^{N/n, N/n, N/n}
  }{\left( 12 C_1 \right)^2} ,
\label{2-160} \\
  \bar \lambda_2 &=& \frac{\Bigl( \tilde \lambda_2 \Bigl)_{n, n, n}^{n, n, n}}
  {\left( 12 C_2 \right)^2} ,
\label{2-170} \\
  \bar \lambda_3 &=& \frac{\Bigl( \tilde \lambda_3 \Bigl)_{n, n, n}^{n, N/n, N/n}}
  {144 C_1 C_2} .
\label{2-175}
\end{eqnarray}
Entities entering the Lagrangian \eqref{2-140} have the following meanings and origins:
\begin{itemize}
	\item the scalar fields $\chi$ and $\phi$ describe the nonperturbatively
  quantized SU(n) and coset SU(N)/SU(n) degrees of freedom, respectively;
	\item the terms $\left( \nabla_\mu \phi \right)^2$ and	
  $\left( \nabla_\mu \chi \right)^2$ are the result of the nonperturbative quantum averaging of
	$(\nabla_\mu A^B_\nu )^2$ in the initial SU(N) Lagrangian;
	\item the terms $\left( \phi^2 - m_1^2 \right)^2$ and
  $[\left( \chi^2 - m_2^2 \right)^2 - m_2^4]$ are the result of the nonperturbative quantum averaging of $f^{ABC} f^{AMN} A^B_\mu A^C_\nu A^{M \mu} A^{N \nu}$;
	\item the term $\phi^2 \chi^2$  is the result of the nonperturbative quantum averaging of $f^{Aab} f^{Amn} A^a_\mu A^b_\nu A^{m \mu} A^{n \nu}$;
  \item the closure coefficients $\lambda_{1,2}$ and $m_{1,2}$ appear;
  \item $C_{1,2}$ are free parameters;
  \item $m_{1,2}$ are eigenvalues
obtained in  solving the field equations \eqref{4-10} and \eqref{4-20}.
\end{itemize}

\section{Field equations}

Using the Lagrangian \eqref{2-140}, one can derive the corresponding field equations describing a gluon condensate in the following form:
\begin{eqnarray}
  \partial_\mu \partial^\mu \phi &=&
  - \phi \left[ \lambda_3 \chi^2 + \lambda_1
  \left(
    \phi^2 - m_1^2
  \right) \right],
\label{4-10}\\
  \partial_\mu \partial^\mu \chi &=&
  - \chi \left[ \lambda_3 \phi^2 + \lambda_2
  \left(
    \chi^2 - m_2^2
  \right) \right],
\label{4-20}
\end{eqnarray}
where
$\phi=\sqrt{N^2-n^2}\bar \phi$,
$\chi=\sqrt{n^2-1}\bar \chi$,
$m_1=\sqrt{N^2-n^2}\bar m_1$,
$m_2=\sqrt{n^2-1}\bar m_2$,
$\lambda_1 = \bar \lambda_1/(N^2 - n^2)^2$,
$ \lambda_2 = \bar \lambda_2/(n^2 - 1)^2$,
$\lambda_3 = \bar \lambda_3/[(N^2 - n^2)(n^2-1)]$.
The coefficients $\lambda_{1,2,3}$ depend on the dimensions $n = \dim SU(n)$ and $N = \dim SU(N)$, where $SU(n) \subset SU(N)$.

Now let us consider one special case.

\subsection{$SU(2) \subset SU(3)$ glueball}

In this case we have $n = 2, N = 3$. The effective Lagrangian will then be
\begin{equation}\label{4-1-10}
  \mathcal L_{eff} = \frac{1}{2}
  \partial_\mu \phi \partial^\mu \phi +
  \frac{1}{2}
  \partial_\mu \chi \partial^\mu \chi -
  \frac{\lambda_1}{4}
   \left( \phi^2 - m_1^2 \right)^2 -
  \frac{\lambda_2}{4}
  \left[
    \left( \chi^2 - m_2^2 \right)^2 - m_2^4
  \right] -
  \frac{\lambda_3}{2}
  \phi^2 \chi^2,
\end{equation}
where $\phi = \tilde \phi \sqrt{60C_1}$, $\chi = \tilde \chi \sqrt{36 C_2}$,
$\lambda_1 = \left( \Delta C_\phi \right)^2/\left( 600 C_1^2 \right)$,
$\lambda_2 = \left( \Delta C_\chi \right)^2/\left( 108 C_2^2 \right)$, and
$\lambda_3 = \Delta C_\phi \Delta C_\chi / \left( 1080 C_1 C_2 \right)$;
$\left| \Delta C^{mn} \right| \approx \Delta C_\phi$ for all $m,n$;
$\left| \Delta C^{ab} \right| \approx \Delta C_\chi$ for all $a,b$. If we assume that $\Delta C_\phi \approx \Delta C_\chi = \Delta C$ and $C_1 = C_2 = C$ then
\begin{equation}\label{4-1-20}
  \lambda_1 = 9 \lambda/5, \quad \lambda_2 = \lambda / 10, \quad
  \lambda_3 = \lambda = (\Delta C)^2/(1080 C^2) .
\end{equation}
For such a case the field equations are
\begin{eqnarray}
  \partial_\mu \partial^\mu \phi &=&
  - \phi \left[ \lambda_3 \chi^2 + \lambda_1
  \left(
    \phi^2 - m_1^2
  \right) \right],
\label{4-1-30}\\
  \partial_\mu \partial^\mu \chi &=&
  - \chi \left[ \lambda_3 \phi^2 + \lambda_2
  \left(
    \chi^2 - m_2^2
  \right) \right]
\label{4-1-40}
\end{eqnarray}
with $\lambda_{1,2,3}$ from \eqref{4-1-20}.

We seek a glueball solution as a spherically symmetric solution with $\phi(r), \chi(r)$. In this case we have the following ordinary differential equations:
\begin{eqnarray}
\phi^{\prime\prime} +\frac{2}{r}\phi^\prime&=&
   \phi \left[ \lambda_3 \chi^2 + \lambda_1
  \left(
    \phi^2 - m_1^2
  \right) \right],
\label{4-1-31}\\
  \chi^{\prime\prime} +\frac{2}{r}\chi^\prime  &=&
   \chi \left[ \lambda_3 \phi^2 + \lambda_2
  \left(
    \chi^2 - m_2^2
  \right) \right],
\label{4-1-41}
\end{eqnarray}
where the prime denotes differentiation with respect to $r$.
Notice that here one can eliminate $\lambda_3$ by redefining
the radial coordinate.
Then, taking into account \eqref{4-1-20}, the system \eqref{4-1-31} and \eqref{4-1-41} takes the form
\begin{eqnarray}
\phi^{\prime\prime} +\frac{2}{x}\phi^\prime&=&
   \phi \left[  \chi^2 + \frac{9}{5}
  \left(
    \phi^2 - m_1^2
  \right) \right],
\label{4-1-32}\\
  \chi^{\prime\prime} +\frac{2}{x}\chi^\prime  &=&
   \chi \left[  \phi^2 + \frac{1}{10}
  \left(
    \chi^2 - m_2^2
  \right) \right],
\label{4-1-42}
\end{eqnarray}
where $x=\sqrt{\lambda} r$ and the prime denotes now differentiation with respect to $x$.

These equations are to be solved subject to the boundary conditions given in the neighborhood of the center by the following expansions:
\begin{equation}
\label{bound_cond}
\phi \approx \phi_0+\frac{1}{2}\phi_2 x^2, \quad \chi \approx \chi_0+\frac{1}{2}\chi_2 x^2,
\end{equation}
where $\phi_0, \chi_0$ are central values of the scalar fields and
the expansion coefficients $\phi_2, \chi_2$ are determined from Eqs.~\eqref{4-1-32} and \eqref{4-1-42}.

For given values of $\phi_0, \chi_0$,
the system of equations ~\eqref{4-1-32} and \eqref{4-1-42} has regular solutions
only for certain values of the masses
of the scalar fields $m_1, m_2$.
As a result, the problem reduces to a search for {\it eigenvalues} of the parameters $m_1, m_2$  and  for the
corresponding {\it eigenfunctions} $\phi$ and $\chi$ of the nonlinear system of
differential equations~\eqref{4-1-32} and \eqref{4-1-42}.
We will seek the specified eigenvalues by using the shooting method. A step-by-step description of the  procedure for finding solutions
can be found, e.g., in Ref.~\cite{shoot}.

\begin{figure}[t]
\centering
  \includegraphics[height=6.5cm]{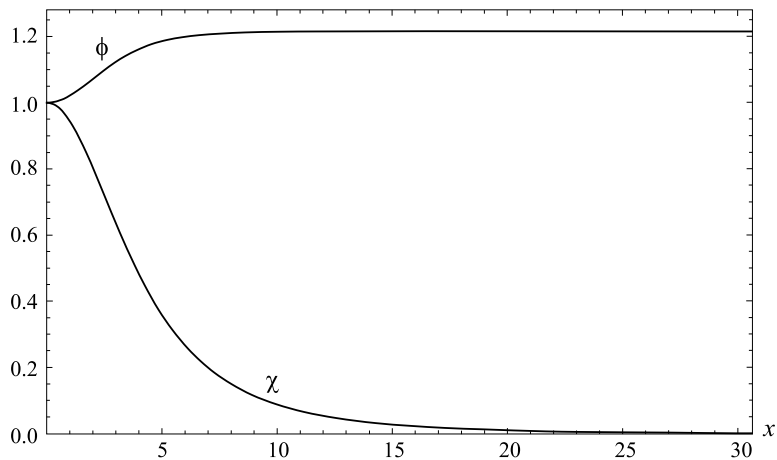}
\vspace{-.7cm}
\caption{The typical behavior of the scalar fields.
 Asymptotically, as $x\to \infty$, the field $\phi$ tends to $m_1$  and
$\chi$ goes to 0 [see  Eqs.~\eqref{asymptotic}-\eqref{asymp_pert}].
}
\label{fig_scal_fields}
\end{figure}

\begin{figure}[h!]
\centering
  \includegraphics[height=6.cm]{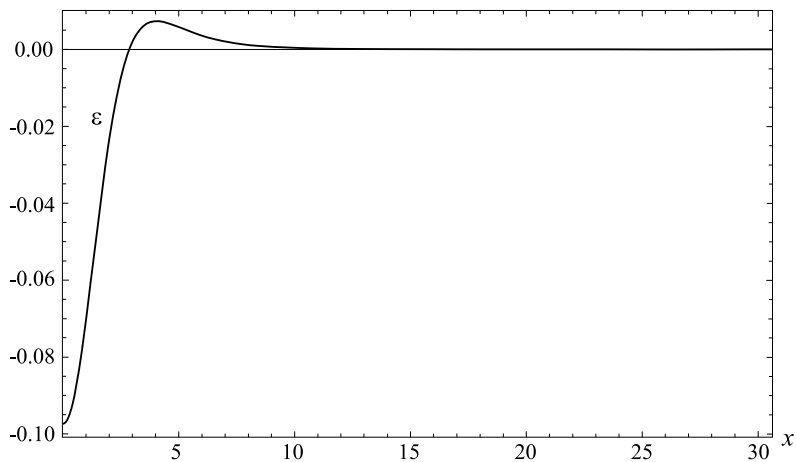}
\caption{The scalar fields  energy density $\varepsilon$ from \eqref{sf_energ_dens}.
}
\label{fig_scal_fields_energy}
\end{figure}

Proceeding in this way, we have obtained the results shown in Figs.~\ref{fig_scal_fields} and \ref{fig_scal_fields_energy}.
It is seen that
$\phi \rightarrow m_1$ and $\chi \rightarrow 0$ at large distances,
i.e., the solutions approach asymptotically the local minimum of the potential energy from the Lagrangian~\eqref{4-1-10}.
One can also see from Fig.~\ref{fig_scal_fields_energy} that the
scalar fields energy density
\begin{equation}
\label{sf_energ_dens}
\varepsilon=\frac{1}{2}\phi^{\prime 2}+\frac{1}{2}\chi^{\prime 2}+
\frac{\lambda_1}{4}\left( \phi^2 - m_1^2 \right)^2 +
  \frac{\lambda_2}{4}
  \left[
    \left( \chi^2 - m_2^2 \right)^2 - m_2^4
  \right] +
  \frac{\lambda_3}{2}
  \phi^2 \chi^2
\end{equation}
goes to zero as $x \rightarrow \infty$.

The asymptotic behavior of the scalar fields can be found in the following form:
\begin{equation}
\label{asymptotic}
    \phi=m_1-\delta \phi, \quad \chi=\delta \chi,
\end{equation}
where
$\delta \phi, \delta \chi \ll 1$ behave as
\begin{equation}
    \delta \phi \approx  k_{\varphi} \frac{\exp{\left(- 3\sqrt{2 m_1^2/5} \,\, x \right)}}{x}, \quad
    \delta \chi \approx  k_{\chi}\frac{\exp{\left(- \sqrt{  m_1^2-m_2^2/10} \,\,x\right)}}{x},
\label{asymp_pert}
\end{equation}
where $k_{\varphi}, k_{\chi}$ are integration constants, and the values of the masses are $m_1\approx 1.216$ and $m_2\approx 3.809$
for the case of $\phi_0=\chi_0=1$ shown in Figs.~\ref{fig_scal_fields} and \ref{fig_scal_fields_energy}.

Using Eq.~\eqref{sf_energ_dens}, one can also find the mass of the glueball:
\begin{equation}
\label{gb_mass}
M=4\pi \int_0^\infty \varepsilon r^2 dr.
\end{equation}
Taking into account that $r=\lambda^{-1/2} x$,  this expression gives the following numerical value: $M\approx 7.4/ \sqrt{\lambda}$.
Using \eqref{4-1-20}, we have for the coupling constant
$\lambda=\left(\Delta C_\phi\right)^2/\left(1080 C_1^2\right)=\left(\Delta C_\chi\right)^2/\left(1080 C_2^2\right)$.

Another possible type of solutions of the system~\eqref{4-1-32}-\eqref{4-1-42} could be solutions for which $\phi\to 0$ and $\chi \to m_2$
asymptotically. However, we have not been successful in obtaining such type of solutions that perhaps indicates their absence.

\section{Conclusion and further problems}

Here we have investigated the scalar model of a glueball proceeding from the nonperturbative quantization ideas \`{a} la Heisenberg. Basic features of the model are:
\begin{itemize}
  \item 2- and 4-point Green functions of quantum fields are described by two scalar fields $\phi, \chi$.
  \item The quantum behavior of gauge fields belonging to the subgroup
  $SU(n) \subset SU(N)$ and $SU(N)$ is different.
  The scalar field $\chi$ describes quantum fluctuations of $A^a_\mu \in SU(n)$, and $\phi$ describes quantum fluctuations of $A^m_\mu \in SU(N)/SU(n)$.
  \item The color space is anisotropic in the sense that
  $\left\langle A^B_{\ldots} A^C_{\ldots} \right\rangle \neq \left\langle A^C_{\ldots} A^B_{\ldots} \right\rangle$ and similarly for 4-point Green functions.
  \item The anisotropy
    $\left\langle A^b_{\ldots} A^b_{\ldots} -  A^c_{\ldots} A^b_{\ldots} \right\rangle = C^{bc} \approx C_{\chi}$ and
      $\left\langle A^m_{\ldots} A^n_{\ldots} -  A^n_{\ldots} A^m_{\ldots} \right\rangle = C^{mn} \approx C_{\phi}$. The constants $C_{\phi, \chi}$ do not depend on color indices.
  \item The dispersion
  $\left\langle \left( A^{B}_{\ldots} \right)^2 \right\rangle$ does not depend on the color index $B$, but may be different for $SU(n)$ and coset $SU(N)/SU(n)$.
  \item 4-point Green functions are some bilinear combinations of 2-point Green functions.
  \item For every group $SU(N)$ there exist different glueballs with different subgroups $SU(n) \subset SU(N)$.
\end{itemize}
There are a few further problems which could be addressed within the framework of the model under discussion:
\begin{itemize}
  \item The comparison with lattice calculations.
  \item The toy model with $b_{\mu \nu} = - \eta_{\mu \nu}$ has negative norm $\left\langle A^{\ldots}_\mu A^{\ldots}_\nu \right\rangle$ for $\mu, \nu = 0$.
  Consideration of more plausible model will be much more complicated.
  \item Study of the asymptotic behavior of the coefficients $\lambda_{1,2,3}$.
  \item Study of the existence of solutions to the field equations with different $\lambda$.
\end{itemize}

\section*{Acknowledgments}

We gratefully acknowledge the support provided
by a grant $\Phi.0755$  in fundamental research in natural sciences by the Ministry of Education and Science of Kazakhstan. We are also grateful to Marco Panero for initiating this work.

\end{document}